\documentclass{aipproc}

   \def\@oddfoot{\reset@font
                 \copyright{} 1998 American Institute of Physics
                 \hfil\@title
                 \hfil\@date\hfil\thepage}
\makeatother

\newcommand{\be}{\begin{equation}}
\newcommand{\ee}{\end{equation}}
\newcommand{\ba}{\begin{eqnarray}}
\newcommand{\ea}{\end{eqnarray}}
\newcommand{\gtwid}{\mathrel{\raise.3ex\hbox{$>$\kern-.75em\lower1ex
\hbox{$\sim$}}}}
\newcommand{\ltwid}{\mathrel{\raise.3ex\hbox{$<$\kern-.75em\lower1ex
\hbox{$\sim$}}}}

\begin{document}

\author{Irina Mocioiu$^{(a)}$ and Robert Shrock$^{(a,b)}$}
\title{Matter Effects on Long Baseline Neutrino Oscillation Experiments}
\affiliation{(a) \ C.N.Yang Institute for Theoretical Physics\\
State University of New York, Stony Brook, NY 11794 \\
(b) Physics Department, Brookhaven National Laboratory \\
Upton, NY  11973 }


\begin{abstract}

We calculate matter effects on neutrino oscillations relevant for long baseline
neutrino oscillation experiments.  In particular, we compare the results
obtained with simplifying approximations for the density profile in the Earth
versus results obtained with actual density profiles.  We study the dependence
of the oscillation signals on both $E/\Delta m^2_{atm.}$ and on the angles in
the leptonic mixing matrix.  The results show quantitatively how matter effects
can cause significant changes in the oscillation signals, relative to vacuum
oscillations and can be useful in amplifying these signals and helping one to
obtain measurements of mixing parameters and the magnitude and sign of
$\Delta m^2$.

\end{abstract}



\section{Introduction}

In a modern theoretical context, one generally expects nonzero neutrino masses
and associated lepton mixing.  Experimentally, there has been accumulating
evidence for such masses and mixing.  All solar neutrino experiments
(Homestake, Kamiokande, SuperKamiokande, SAGE, and GALLEX) show a significant
deficit in the neutrino fluxes coming from the Sun \cite{solexp}. This deficit
can be explained by oscillations of the $\nu_e$'s into other weak
eigenstate(s), with $\Delta m^2_{sol}$ of the order $10^{-5}$ eV$^2$ for MSW
solutions \cite{msw} or of the order of $10^{-10}$ eV$^2$ for vacuum
oscillations.  Accounting for the data with vacuum oscillations requires almost
maximal mixing.  The MSW solutions include one for small mixing angle (SMA) and
one with essentially maximal mixing (LMA).

Another piece of evidence for neutrino oscillations is the atmospheric neutrino
anomaly, observed by Kamiokande, SuperKamiokande, IMB, MACRO, and Soudan-2
\cite{atmos}.  Of these, the Superkamiokande data has especially high
statistics - roughly 52 kton-years worth of data at present.  This data can be
well fit by the inference of $\nu_{\mu} \rightarrow \nu_x$ oscillations with
$\Delta m^2_{atm}\sim 3.5 \times 10^{-3}$ eV$ ^2$ \cite{atmos} and maximal mixing
$\sin^2 2 \theta_{atm} = 1$, where $\nu_x = \nu_\tau$ is favored.  The
possibility $\nu_x = \nu_{sterile}$ is disfavored at the $2\sigma$ level
\cite{learned}.  (The possibility that $\nu_x$ is predominantly $\nu_e$ is
ruled out by both the Superkamiokande data and the CHOOZ experiment
\cite{chooz}).
  
In addition, the LSND experiment has reported observing $\bar\nu_\mu \to 
\bar \nu_e$ and $\nu_{\mu} \to \nu_e$ oscillations with $\Delta m^2_{LSND} \sim
0.1 - 1$ eV$^2$ and moderately small mixing angle. This result is not 
confirmed by a similar experiment, KARMEN \cite{lsnd}.  

There are currently strong efforts to confirm and extend the evidence for
neutrino oscillations in all of the various sectors -- solar, atmospheric, and
accelerator.  Some of these are currently running: the Sudbury Neutrino
Observatory, SNO, the K2K pioneering long baseline experiment between KEK and
Kamioka.  Others are in development and testing phases, such as Borexino,
KamLAND, MINOS, mini-BOONE, and the CERN-Gran Sasso program.  Among the long
baseline neutrino oscillation (LBLNO) experiments, the distances are $L \simeq
250$ km for K2K, 730 km for both MINOS, from Fermilab to Soudan and the
proposed CERN-Gran Sasso experiments.  The sensitivity of these experiments
should reach the region $ \Delta m^2 \sim {\rm \,\,\, few} \times 10^{-3}
$eV$^2$.
Another generation of experiments, with even higher sensitivity will be
required for precision measurements of oscillation parameters.  One of the
physics capabilities of the Next generation Nucleon decay and Neutrino detector
discussed at this NNN99 workshop would be as part of a LBLNO experiment.
An interesting possibility that is being studied intensively is a muon collider
or storage ring that would serve as a source of quite high intensity,
flavor-pure ($\nu_\mu + \bar\nu_e$ beams from $\mu^-$ and $\bar\nu_\mu + \nu_e$
beam from $\mu^+$) (anti)neutrino beams.  Using these, one could
perform LBLNO experiments with an existing deep underground detector, e.g., at
Soudan, Gran Sasso, or Kamioka, the NNN detector, and/or a surface detector.
Studies have shown that one can get hundreds of events per kiloton-year at
distances of 7000-9000 km \cite{geer}, \cite{barger}.
It is thus appropriate to begin planning for this next generation of very 
long baseline neutrino oscillation experiments.  

An important effect that must be taken into account in such experiments is the
matter-induced oscillations which neutrinos undergo along their flight path
through the Earth from the source to the detector.  In a hypothetical world in
which there were only two neutrinos, $\nu_\mu$ and $\nu_\tau$, the $\nu_\mu \to
\nu_\tau$ oscillations in matter would be the same as in vacuum, since both
have the same forward scattering amplitude, via $Z$ exchange, with matter.
However, in the realistic case of three generations, because of the indirect
involvement of $\nu_e$ due to a nonzero $U_{13}$, and because of the fact that
$\nu_e$ has a different forward scattering amplitude off of electrons,
involving both $Z$ and $W$ exchange, there will be a matter-induced oscillation
effect on $\nu_\mu \to \nu_\tau$ (as well as other channels).  An early study
of matter effects in the earth is \cite{baltz}; several recent studies are
\cite{barger}-\cite{mlb}.

Here we shall report on a study that we have carried out \cite{ms} of matter
effects relevant to LBLNO experiments.  We consider the usual three flavors of
active neutrinos, with no light sterile (= electroweak-singlet) neutrinos.
This is sufficient to describe the more established evidence from the solar and
atmospheric neutrino deficit (if one tried to fit also the LSND experiment with
a neutrino oscillation scenario, one would be led to include light sterile
neutrinos).  As suggested by the solar and atmospheric data, we consider that
there is only one mass scale relevant for long baseline and atmospheric
neutrino oscillations, $\Delta m^2_{atm} \sim \ {\rm few} \ \times 10^{-3}$
eV$^2$ and we work with the hierarchy 
\be 
\Delta m^2_{21} = \Delta m^2_{sol}
\ll \Delta m^2_{31} \approx \Delta m^2_{32}=\Delta m^2_{atm}
\label{hierarchy}
\ee 

In our work we take into account the actual profile of the Earth, as given by
geophysical seismic data \cite{prem} and compare the results with those
calculated using the approximation of average density along the path of the
neutrino.  Further, we present the oscillation probabilities as functions of
$E/\Delta m^2$ so one can determine which energies are best suited for precise
measurements of $\Delta m^2$ in a given region.  We study how these oscillation
probabilities vary with the different input parameters and discuss the
influence of the matter effects on the sensitivity to each of these parameters.

\section{Matter Effects}

The evolution of the flavor eigenstates is given by
\be
{ i}\frac{ d}{ d x}\nu =\left(\frac{1}{2E}UM^2U^{\dagger}+V\right)\nu
\ee 
where the flavor neutrino wavefunction is 
\be
\nu=U\nu_m
\ee
in terms of the mass eigenstates 
\be
\nu_m=\pmatrix{\nu_1\cr \nu_2\cr\nu_3}
\ee
and 
\be
M^2=\pmatrix{m_1^2&0&0\cr0&m_2^2&0\cr0&0&m_3^2},
V=\pmatrix{\sqrt{2}G_FN_e&0&0\cr0&0&0&\cr0&0&0}
\label{V}
\ee
where $N_e$ is the electron number density and 
$\sqrt{2}G_FN_e [{\rm eV}] = 7.6\times 10^{-14} Y_e \rho$  [g/cm$^3$].

The leptonic mixing matrix U can be written as
\be
U=R_{23}KR_{13}K^*R_{12}K'
\ee  
which is the standard CKM-type parametrization, with $R_{ij}$ being the 
rotation matrix in the $ij$ subspace, $c_{12}=\cos\theta_{12}$, 
$s_{12}=\sin\theta_{12}$, etc., $K=diag(e^{-i\delta},1,1)$, and
$K'=diag(e^{i\alpha_1},e^{i\alpha_2},1)$ (the latter phases originate from
the general presence of Majorana mass terms but will not be important here).  

The atmospheric neutrino data suggests almost maximal mixing in the $2-3$
sector. However, a small but non-zero $s_{13}$ is still allowed, and this
produces the matter effect in the traversal of neutrinos through the Earth.  We
use $\sin^2(2\theta_{13}) \le 0.1$, consistent with the limits from the
atmospheric neutrino data \cite{atmos} and the CHOOZ experiment
\cite{chooz}. We also assume the small mixing angle (SMA) MSW solution to the
solar neutrino data.  This assumption, together with the hierarchy of eq. 
(\ref{hierarchy}), implies that, for the relevant energies $E \gtwid 1$ GeV 
and pathlengths $L \sim 10^3 - 10^4$ km, only one squared mass scale, $\Delta
m^2_{atm}$, is important for the oscillations, and the three-species neutrino
oscillations can be described in terms of this quantity, $\Delta m^2_{atm}$,
and the mixing parameters $\sin^2(2\theta_{23})$, and $\sin^2(2\theta_{13})$, 
with negligible dependence on $\sin^2(2\theta_{12})$ and $\delta$; hence also, 
CP violation effects would be negligibly small here, and $P(\nu_a \to \nu_b)
= P(\nu_b \to \nu_a)$, $P(\bar\nu_a \to \bar\nu_b) = 
P(\bar\nu_b \to \bar\nu_a)$.  Although, {\it a priori}, CP violation would 
lead to $P(\nu_a \to \nu_b) \ne P(\bar\nu_a \to \bar\nu_b)$ in vacuum, this
inequality is true in matter even in the absence of CP violation. 

\begin{figure}
 \resizebox{1.\columnwidth}{!}
  {\includegraphics[draft=false]{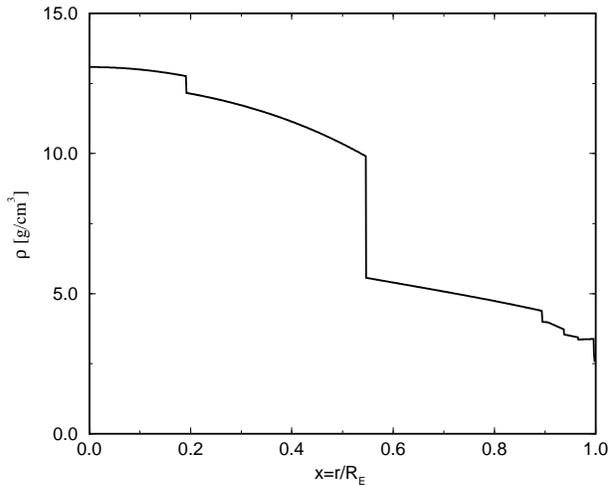}}
 \caption{Density profile of the Earth}
 \label{fig:density}
\end{figure}

For our purposes, we recall that the Earth is composed of crust, mantle, liquid
outer core, and solid inner core, together with additional sublayers in the
mantle, with particularly strong changes in density between the lower mantle
and outer core. The density profile of the Earth is shown in fig.
\ref{fig:density} from \cite{prem}.  The core has average density
$\rho_{core}=11.83$ g/cm$^3$ and electron fraction $Y_{e,core}=0.466$, while
the mantle has average density $\rho_{mantle}=4.66$ g/cm$^3$ and
$Y_{e,mantle}=0.494$.  If one approximates the density as a constant along the
neutrino flight path, the evolution equation can easily be solved, with
well-known results.  However, when one takes account of the actual
variable-density situation in the earth, it is necessary to perform a numerical
integration of the evolution equation, which we have done.

\section{Results and Discussion}

  For long baseline experiments like K2K, Fermilab to Minos, and CERN to
Gran-Sasso, the neutrino flight path only goes through the upper mantle.  The
density in this region is practically constant, and the oscillation
probabilities can easily be calculated. The matter effects are small, but
possibly detectable for the longer baselines.  However, there are several
motivations for very long baseline experiments, since, with sufficiently
high-intensity sources, these can be sensitive to quite small values of $\Delta
m^2$ and since the matter effects, being larger, can amplify certain
oscillations and can, in principle, be used to get information on the sign of
$\Delta m^2_{atm}$.  Hence we concentrate here on these very long baseline
experiments; for these, the neutrino flight path goes through several layers of
the Earth with different densities, including the lower mantle.  We show
results for $L\simeq 7330$ km, the distance from Fermilab to Gran Sasso.  We
have also performed calculations for the Fermilab to SuperKamiokande and
Fermilab to SLAC path lengths, $\sim$ 9120 and 2880 km, respectively.  We
calculate the probabilities of oscillation in long baseline experiments as a
function of $E/\Delta m^2$, rather than using a particular value for $\Delta
m^2$ or the energy.  The relevant ranges are $\Delta m^2 \sim \ {\rm few}
\times 10^{-3}$ eV$^2$ and energies $E$ of the order of tens of GeV. This way
of presenting the results can be useful in studying the optimization of the
beam energy.  In our work \cite{ms} we calcualte the oscillation probabilities
for different values of the mixing angles $\theta_{13}$ and $\theta_{23}$
allowed by the atmospheric neutrino data and the CHOOZ experiment; for this
workshop report we only show results for $\sin^2(2\theta_{23})=1$.  We consider
both neutrinos and antineutrinos. The matter effects change sign in these two
cases; for antineutrinos, $V$ in (\ref{V}) is replaced with $(-V)$.  This
implies that if $\Delta m^2$ is positive (as considered here), one can get a
resonant enhancement of the oscillations for neutrinos,
while for antineutrinos the matter effects would suppress the oscillations.
The situation would be reversed if $\Delta m^2$ were negative. 

\begin{figure}
 \resizebox{1.\columnwidth}{!}
  {\includegraphics[draft=false]{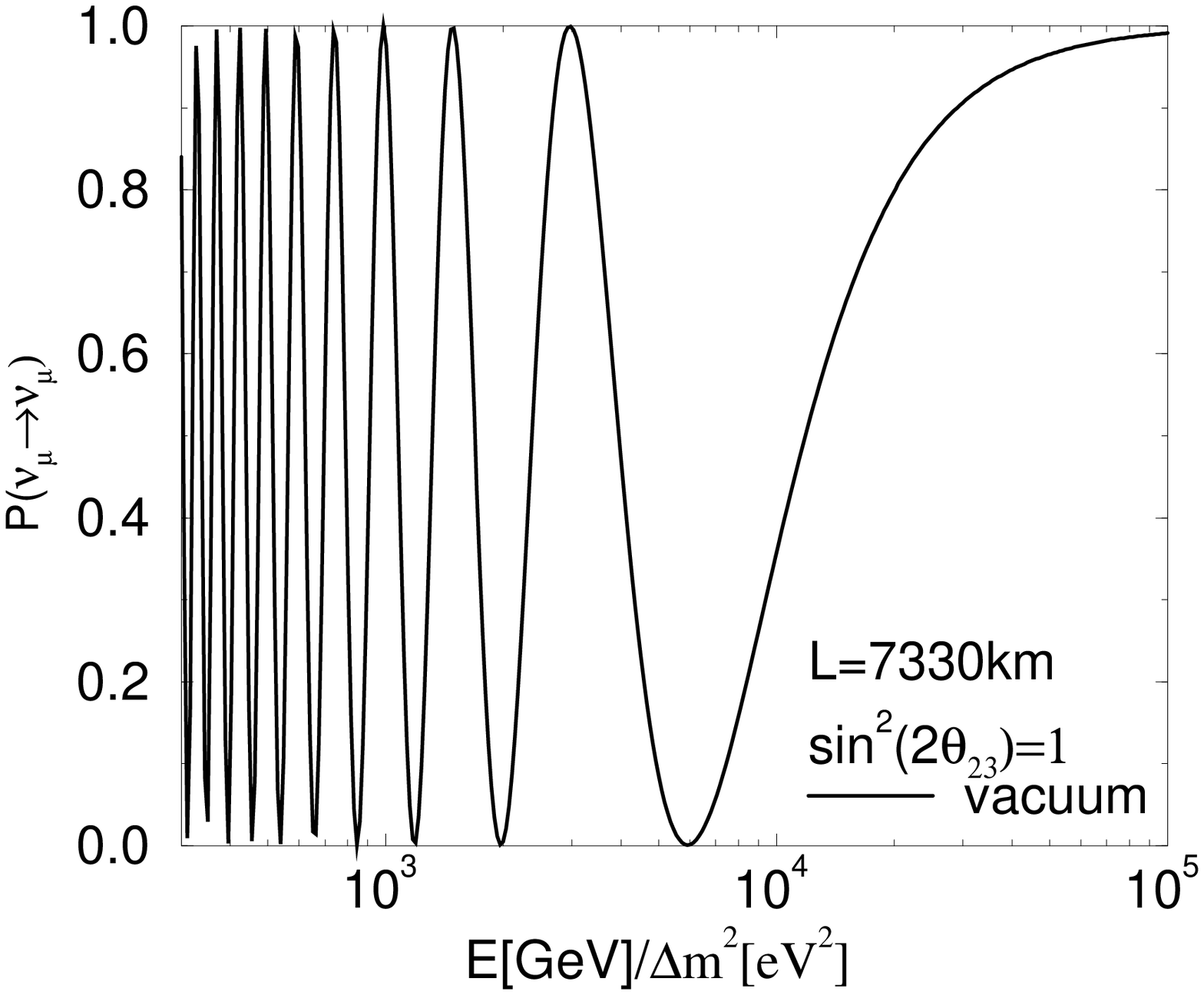}}
 \caption{}
 \label{fig:fgsvm11}
\end{figure}

\begin{figure}
 \resizebox{1.\columnwidth}{!}
  {\includegraphics[draft=false]{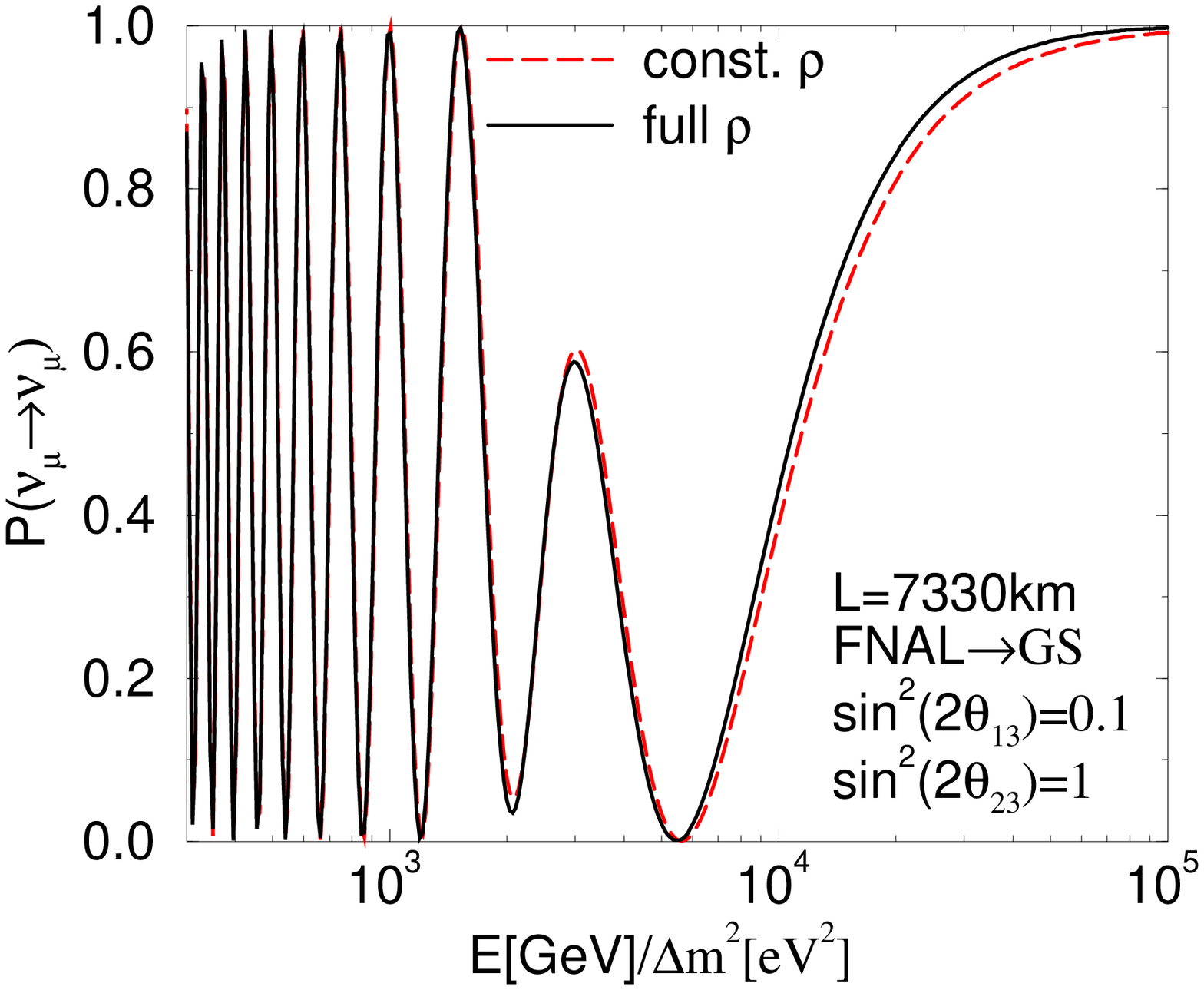}}
 \caption{}
 \label{fig:fgsm11}
\end{figure}

\begin{figure}
 \resizebox{1.\columnwidth}{!}
  {\includegraphics[draft=false]{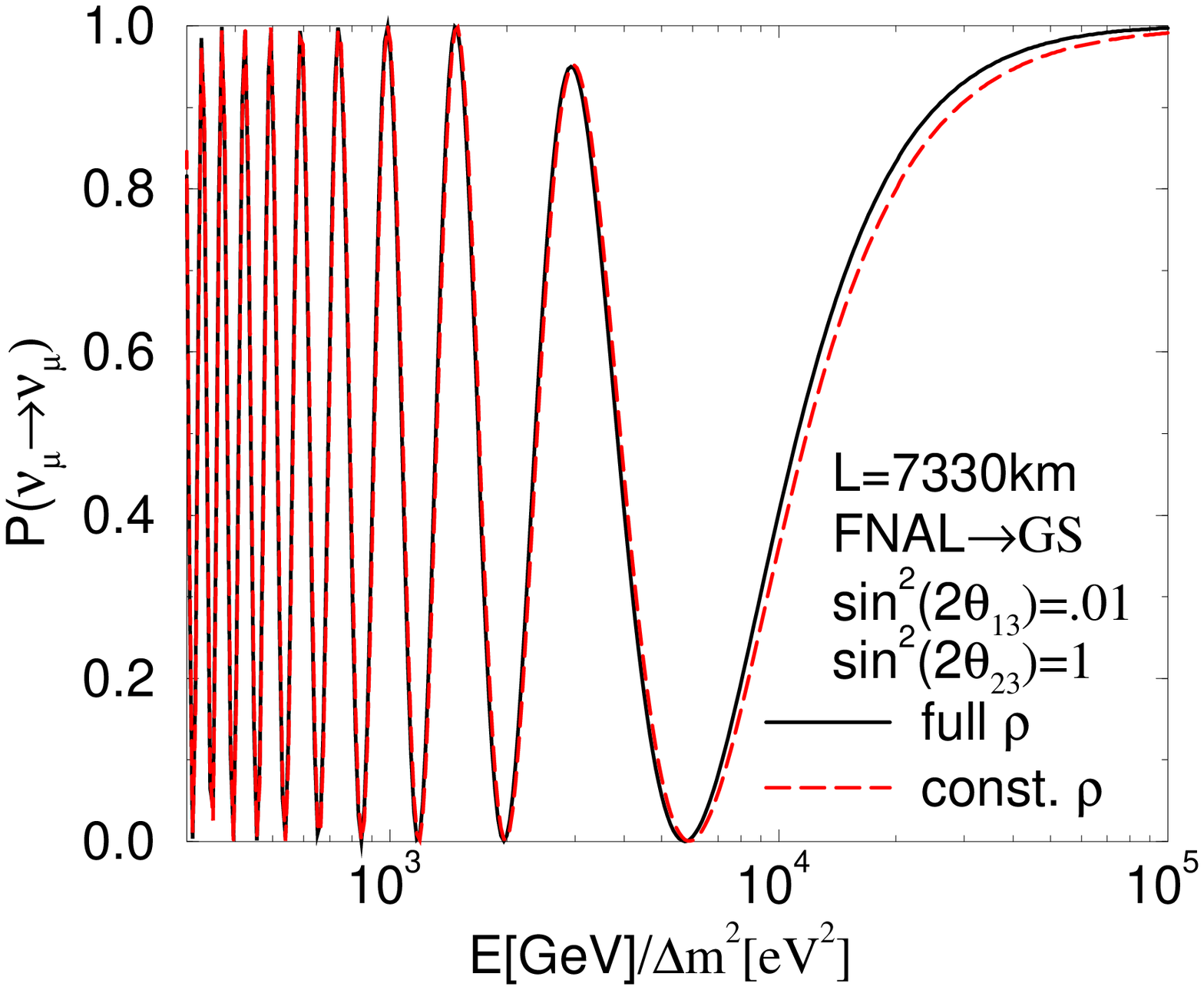}}
 \caption{}
 \label{fig:fgsm011}
\end{figure}

We first study the survival probability of $\nu_\mu$. If the beam went 
through vacuum, the oscillation probability would look like the curve in fig.
\ref{fig:fgsvm11} for practically any value of $\sin^2(2\theta_{13})$. 
In matter, this probability becomes sensitive to all oscillation parameters, as
can be seen from fig. \ref{fig:fgsm11} and fig. \ref{fig:fgsm011}.

\begin{figure}
 \resizebox{1.\columnwidth}{!}
  {\includegraphics[draft=false]{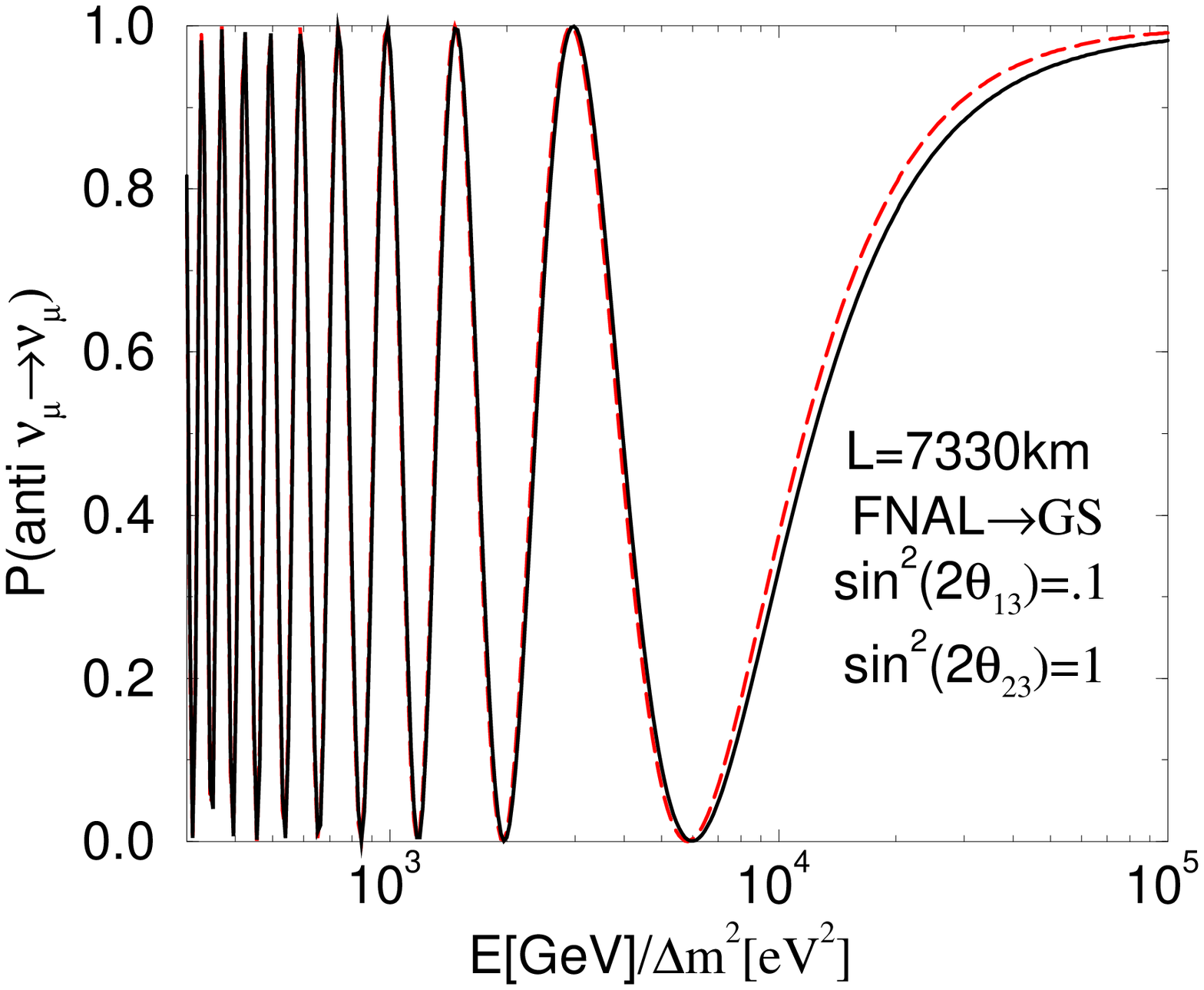}}
 \caption{}
 \label{fig:afgsm11}
\end{figure}

We also want to compare the solution in vacuum (fig. \ref{fig:fgsvm11}) 
with the solution in matter for neutrinos (fig. \ref{fig:fgsm11}) and 
antineutrinos (fig. \ref{fig:afgsm11}).  In the legends for the figures with
antineutrinos, ``anti $\nu_a \to \nu_b$ means $\bar\nu_a \to 
\bar\nu_b$.  One can see the opposite effects of matter on
neutrinos and antineutrinos. The difference in the results for different 
mixing angles makes it possible in principle to use this probability for 
relatively precise measurements of the oscillation parameters.  
Measuring separately the probability  for
$\nu$ and $\bar\nu$ can be very useful in detecting the matter 
effects and using these to constrain the relevant mixings and squared mass 
difference.  Clearly, if one could use two path lengths, as may be possible
with a neutrino factory, this would provide more information and constraints. 

\begin{figure}
 \resizebox{1.\columnwidth}{!}
  {\includegraphics[draft=false]{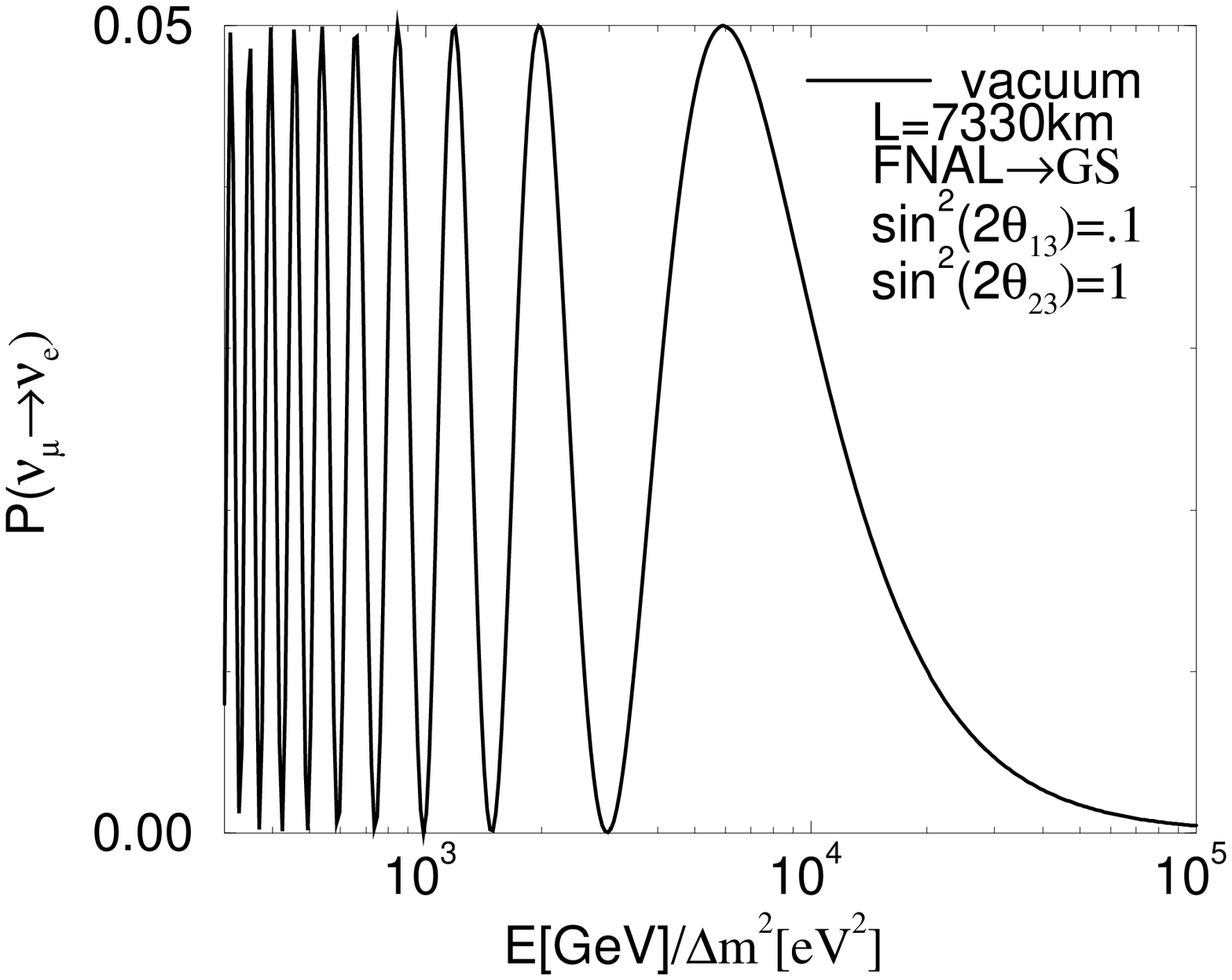}}
 \caption{}
 \label{fig:fgsve11}
\end{figure}

\begin{figure}
 \resizebox{1.\columnwidth}{!}
  {\includegraphics[draft=false]{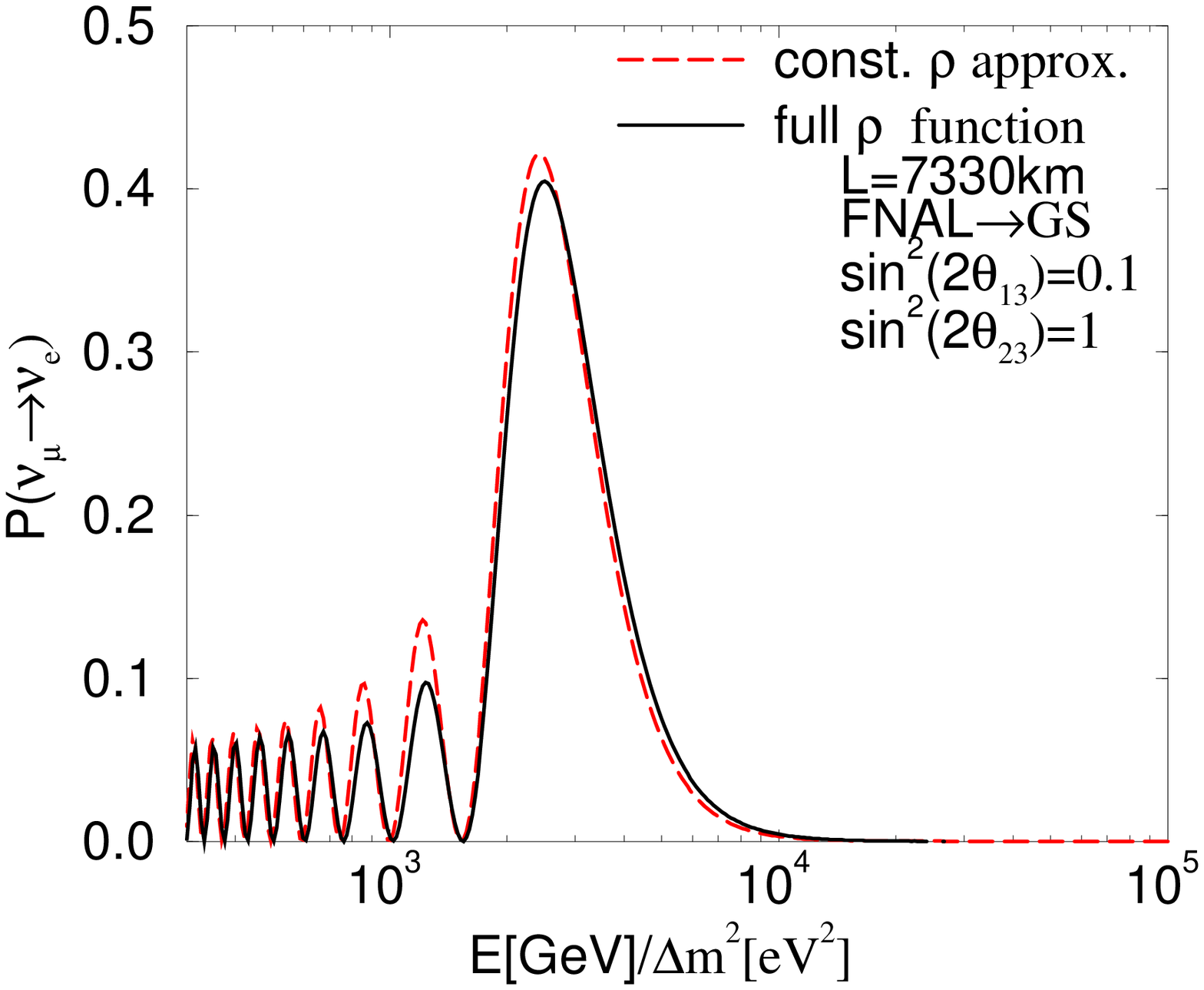}}
 \caption{}
 \label{fig:fgse11}
\end{figure}

\begin{figure}
 \resizebox{1.\columnwidth}{!}
  {\includegraphics[draft=false]{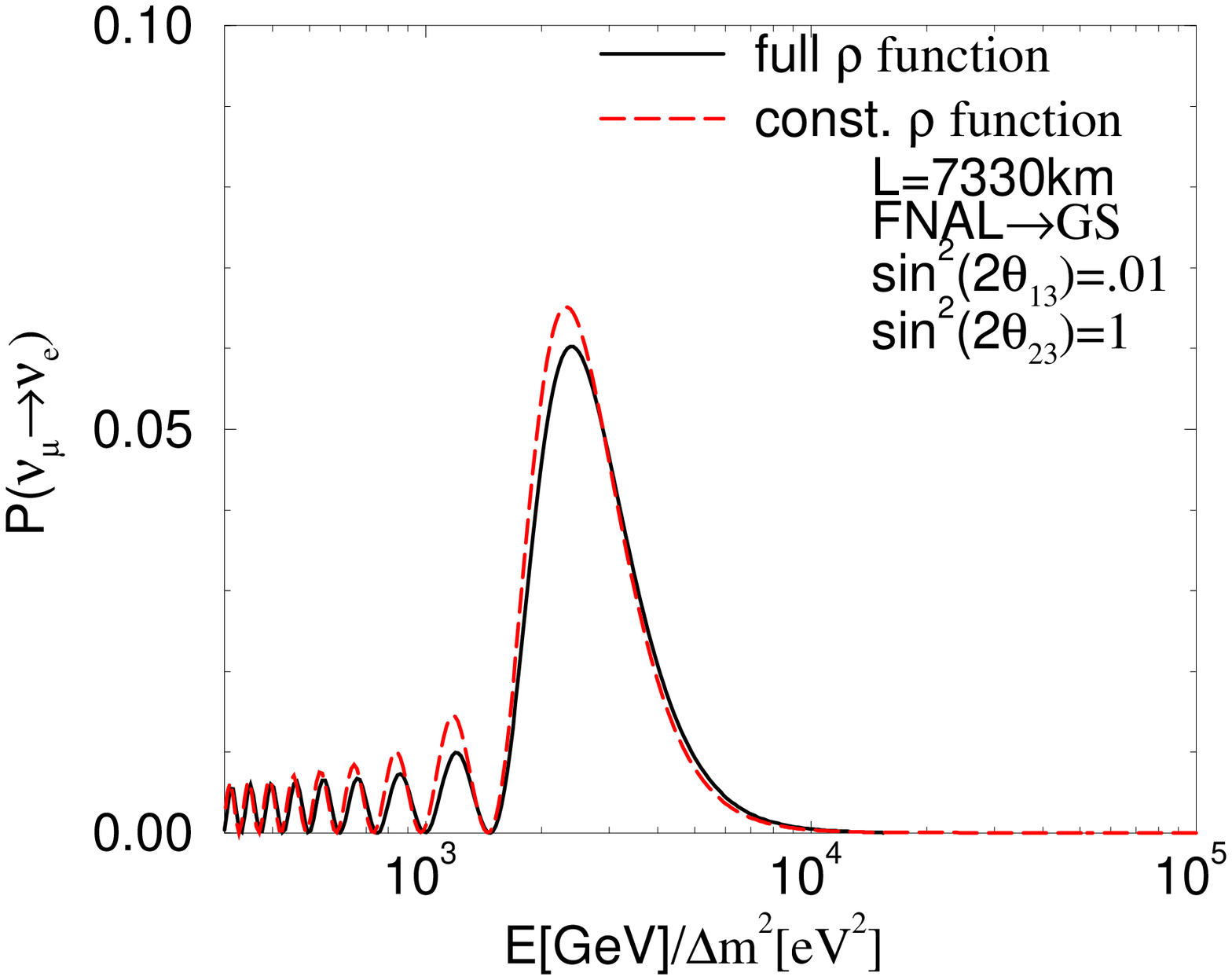}}
 \caption{}
 \label{fig:fgse011}
\end{figure}

\begin{figure}
 \resizebox{1.\columnwidth}{!}
  {\includegraphics[draft=false]{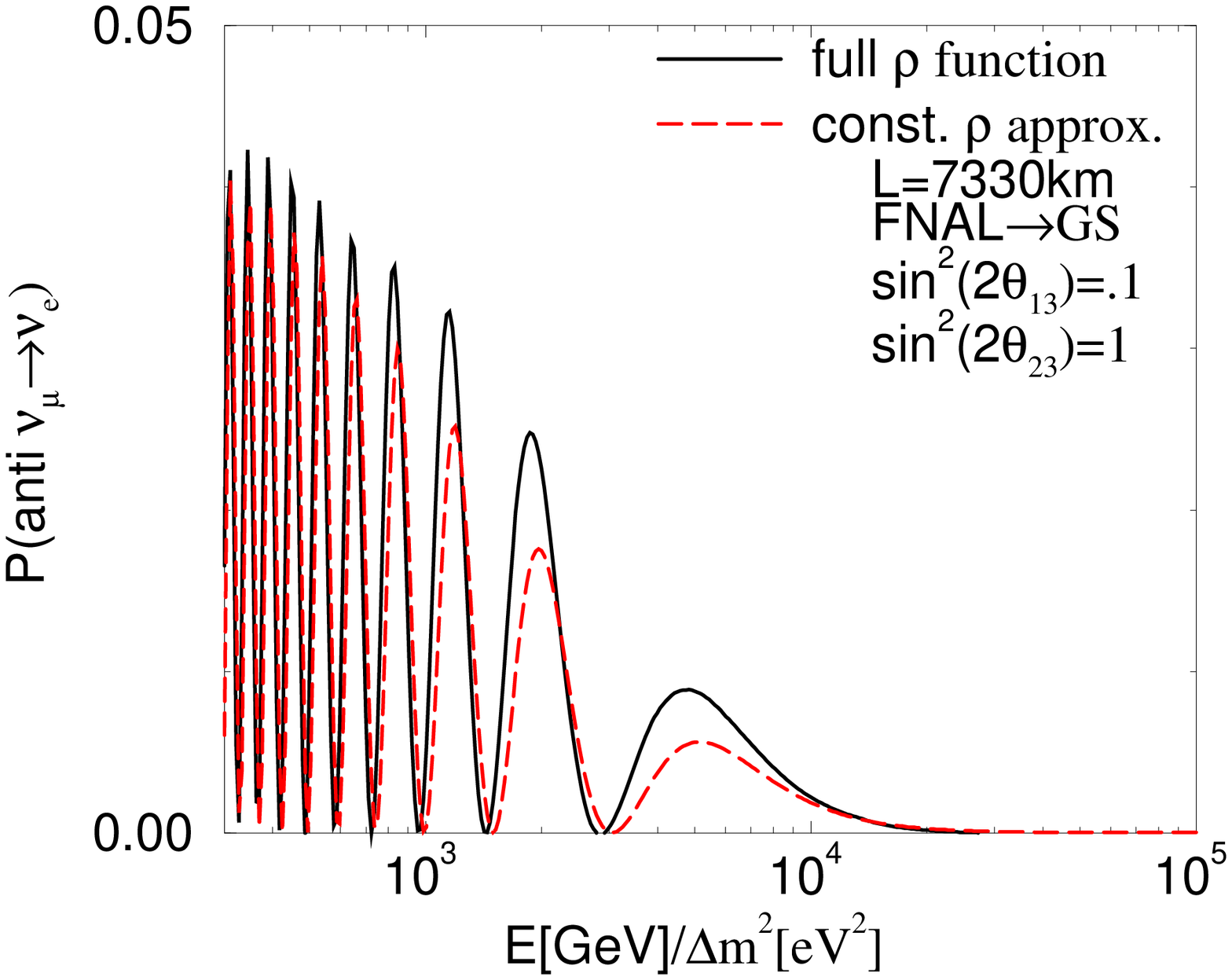}}
 \caption{}
 \label{fig:afgse11}
\end{figure}

The relative effects of matter can be especially dramatic in the oscillation
probability $P(\nu_\mu\rightarrow\nu_e)$, since these directly involve $\nu_e$.
If the beam were to go through the vacuum, $P(\nu_\mu\rightarrow\nu_e)$ would
be probably too small to detect (fig. \ref{fig:fgsve11}).  Because of the
matter effect however, this probability can be strongly enhanced, as is evident
in fig. \ref{fig:fgse11}.  The enhancement is largest for $E/\Delta m^2$ around
3000 GeV/eV$^2$.  This is essentially equal to the ratio that one would get
using a beam energy of $\sim 10$ GeV, given the indication from the data that
$\Delta m_{atm} = 3.5 \times 10^{-3}$ eV$^2$.  Hence the matter effect can
amplify $P(\nu_\mu\rightarrow\nu_e)$ and enable this transition to be measured
with reasonable accuracy, thereby yielding important information on the
oscillation parameters.  This probability is very sensitive to the value of
$\theta_{13}$ (figs. \ref{fig:fgse11}, \ref{fig:fgse011}), so one could use it
for a good determination of this angle.  The sensitivity to $\Delta m^2$ is
also quite strong, due to the pronounced peak given by the matter effect in the
relevant region. Note that for antineutrinos, the oscillation is suppressed
(fig. \ref{fig:afgse11}), so an independent measurement of the two channels
($\nu_\mu\rightarrow \nu_e$ and $\bar\nu_\mu\rightarrow \bar\nu_e$) would be
very valuable.

\begin{figure}
 \resizebox{1.\columnwidth}{!}
  {\includegraphics[draft=false]{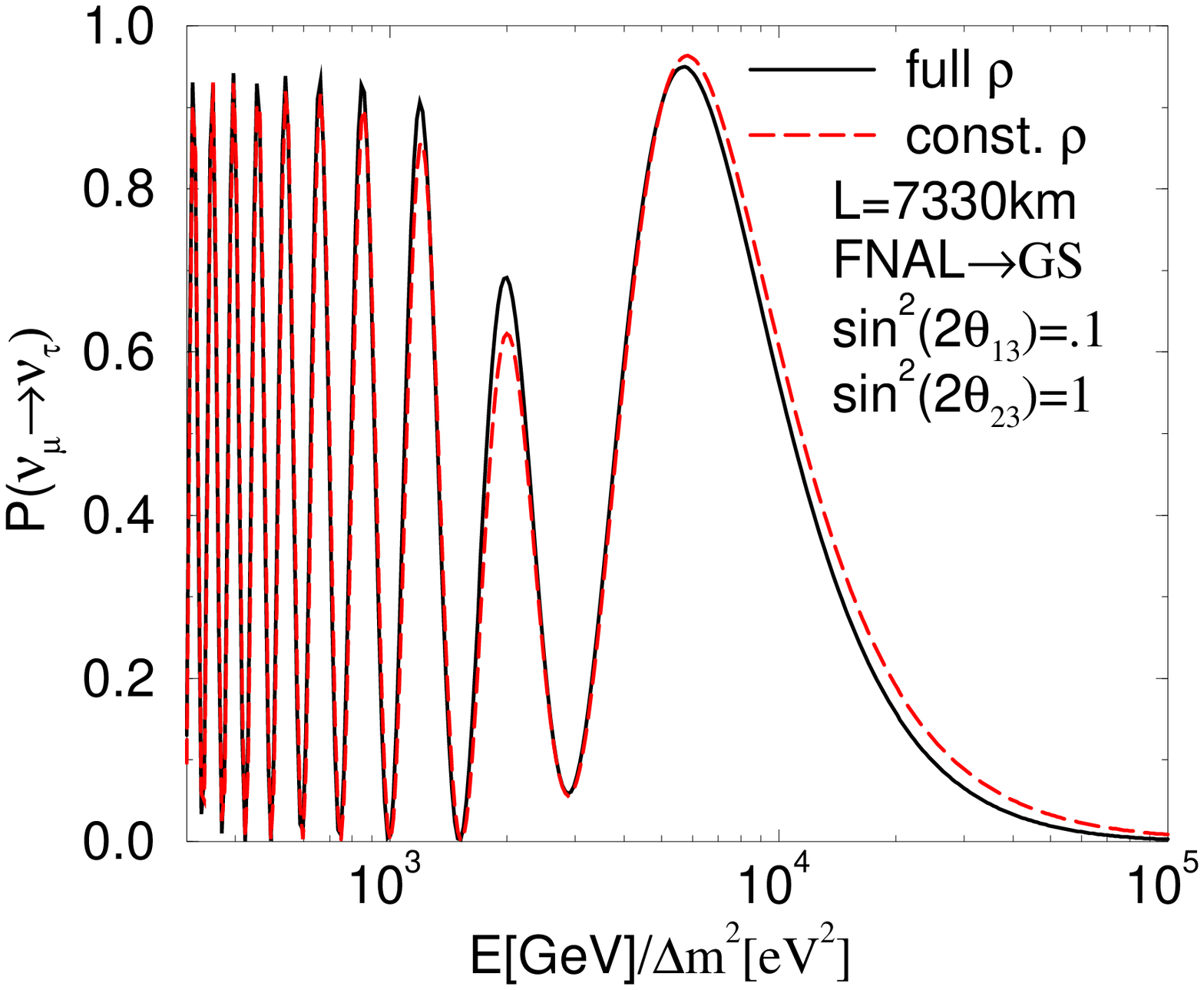}}
 \caption{}
 \label{fig:fgst11}
\end{figure}

\begin{figure}
 \resizebox{1.\columnwidth}{!}
  {\includegraphics[draft=false]{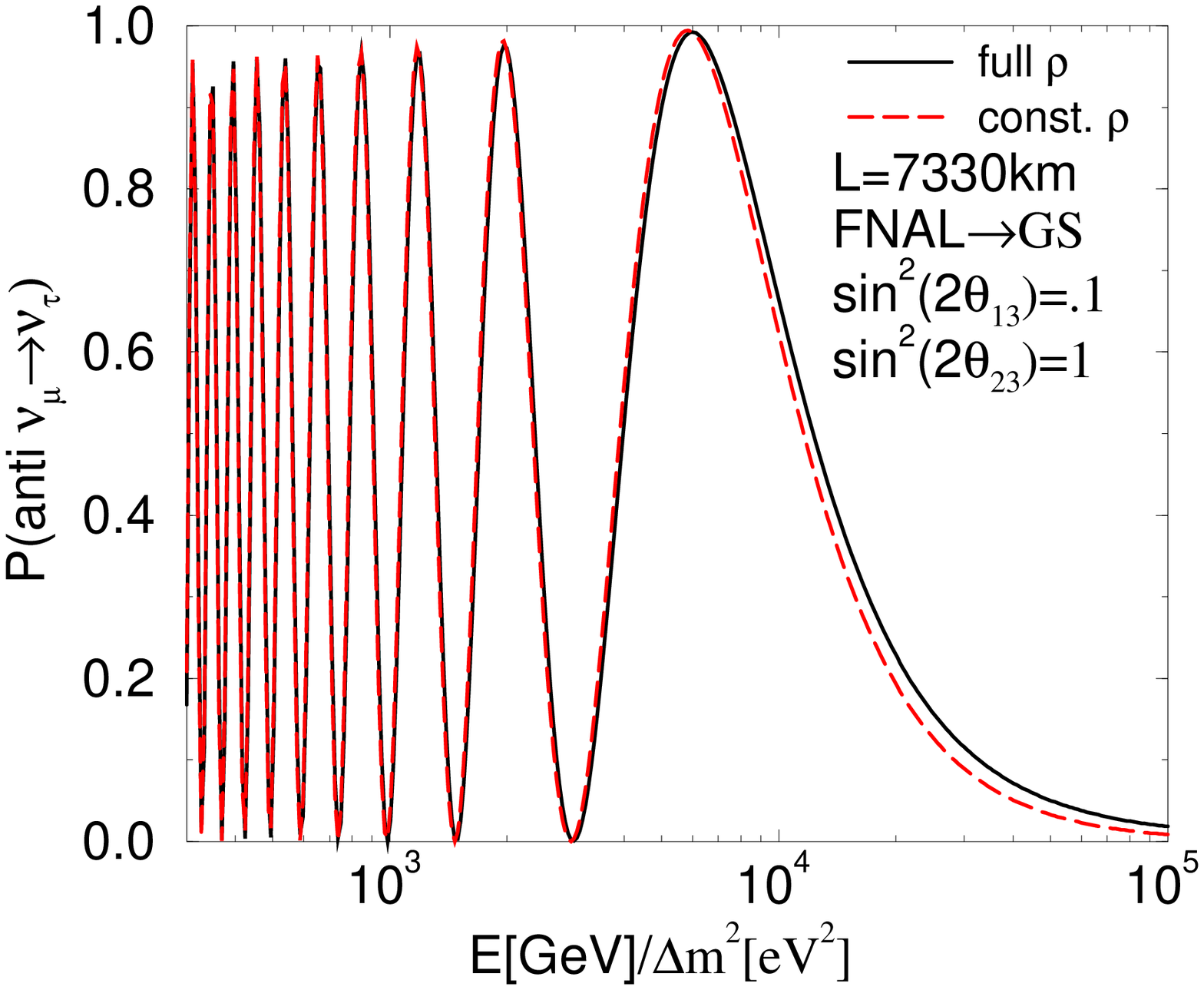}}
 \caption{}
 \label{fig:afgst11}
\end{figure}

The atmospheric neutrino data tells us that the dominant oscillation channel is
actually $\nu_\mu\rightarrow \nu_\tau$.  Consequently, it would be very useful
to measure $P(\nu_\mu\rightarrow\nu_\tau)$; this would provide further
confirmation of this oscillation and could also provide accurate determinations
of $\Delta m^2$ and $\theta_{23}$.  Fig. \ref{fig:fgst11}
shows $P(\nu_\mu\rightarrow\nu_\tau)$. Fig. \ref{fig:afgst11} shows 
$P(\bar\nu_\mu\rightarrow\bar\nu_\tau)$.
 
\begin{figure}
 \resizebox{1.\columnwidth}{!}
  {\includegraphics[draft=false]{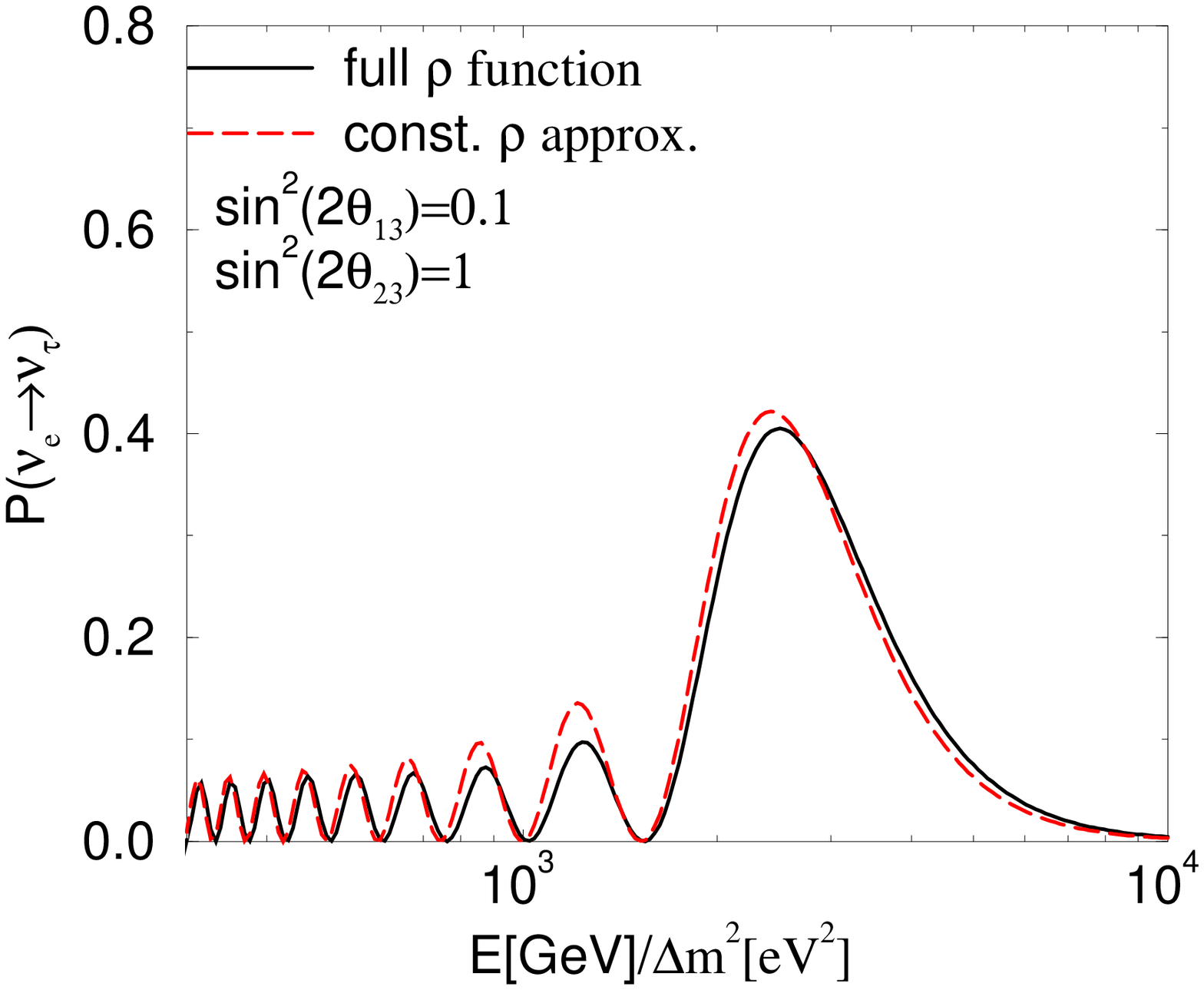}}
 \caption{}
 \label{fig:fgset11}
\end{figure}

\begin{figure}
 \resizebox{1.\columnwidth}{!}
  {\includegraphics[draft=false]{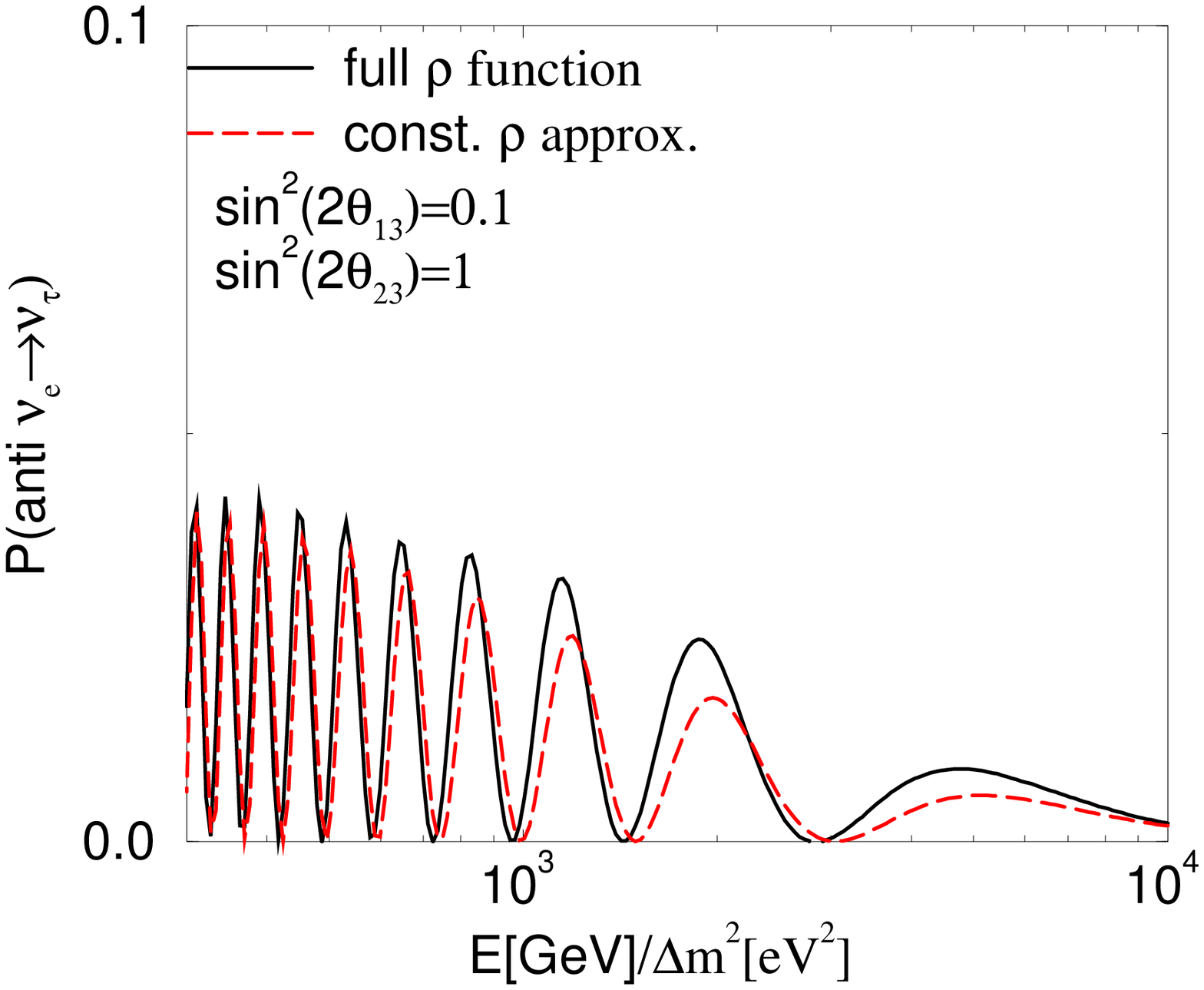}}
 \caption{}
 \label{fig:afgset11}
\end{figure}

Since with a muon collider or muon storage ring, $\nu_e$ ($\bar \nu_e$) beams 
would also be available, it would be interesting to study oscillation 
probabilities with these beams.  We already have the results for 
$P(\nu_e\rightarrow \nu_\mu)$ since, as mentioned above, with our parameters,
this is the same as $P(\nu_\mu\rightarrow \nu_e)$. We present here 
$P(\nu_e\rightarrow \nu_\tau)$ in fig. \ref{fig:fgset11} and 
$P(\bar\nu_e\rightarrow \bar\nu_\tau)$ in Fig. \ref{fig:afgset11}.  
These calculations show that matter effects are important and enhance 
oscillations of the neutrinos and suppress oscillations of antineutrinos in 
the relevant region of parameters.

To summarize, in planning for very long baseline neutrino oscillation
experiments, it is important to take into account matter effects.  We have
performed a careful study of these, including realistic density profiles in the
earth.  Matter effects can be useful in amplifying neutrino oscillation signals
and helping one to obtain measurements of mixing parameters and the magnitude
and sign of $\Delta m^2$.  

\vspace{6mm}

  We thank Debbie Harris for some helpful comments.
  The research of R. S. was supported in part at Stony Brook by the U. S. NSF
grant PHY-97-22101 and at Brookhaven by the U.S. DOE contract
DE-AC02-98CH10886. Accordingly, the U.S. government retains a non-exclusive
royalty-free license to publish or reproduce the published form of this
contribution or to allow others to do so for U.S. government purposes.

\end{document}